\documentclass[aps,twocolumn,pra,superscriptaddress,groupedaddress]{revtex4}

\usepackage{amssymb} 
\usepackage{color,graphicx} \usepackage{amsmath}
\usepackage{mathtools}
\usepackage{amsbsy} \usepackage{amsthm}
\usepackage{float} \usepackage{braket}
\usepackage{placeins}
\usepackage[colorlinks=true,citecolor=blue,linkcolor=red,urlcolor=red]{hyperref}
\usepackage[sort&compress]{natbib}
\usepackage{comment,soul}
\usepackage{stackengine}
\usepackage[inkscapeformat=png]{svg}
\usepackage{array}
\usepackage{physics}
\newcolumntype{C}{>{$\displaystyle} c <{$}}

\newcommand\D{\operatorname{d}\!}

\begin{document}
\author{Oliviero Angeli}
\affiliation{Department of Physics, University of Trieste, Strada Costiera 11, 34151 Trieste, Italy}
\affiliation{Istituto Nazionale di Fisica Nucleare, Trieste Section, Via Valerio 2, 34127 Trieste, Italy}

\author{Matteo Carlesso}
\affiliation{Department of Physics, University of Trieste, Strada Costiera 11, 34151 Trieste, Italy}
\affiliation{Istituto Nazionale di Fisica Nucleare, Trieste Section, Via Valerio 2, 34127 Trieste, Italy}

\title{Entanglement in Markovian Hybrid Classical-Quantum Theories of Gravity}

\date{\today}

\begin{abstract}
The entangling properties of models of classical gravity interacting with quantum matter (i.e.~hybrid models of gravity) are investigated in the context of the experimental proposals to detect gravitationally induced entanglement.  We prove that entanglement generation can indeed take place within these models, and characterize it quantitatively. We identify the root of entanglement generation in these models in the presence of some underlying non-locality of the theories. Nonetheless, by focusing in particular on the Di\'osi-Penrose model for two gravitationally interacting masses, we show that entanglement-based experiments have the potential to either falsify the model entirely or constrain the free parameter of the model $R_0$ up to values six orders of magnitude above the current state of the art.
\end{abstract}
\maketitle

\section{Introduction}
Understanding the interplay between gravitational phenomena and quantum mechanics is undoubtedly one of the greatest challenges of contemporary physics. While vast research programs, like string theory~\cite{Schellekens2013} or loop quantum gravity~\cite{rovelli2008loop,ashtekar2021short}, have attempted to provide a unified framework of quantum and gravitational phenomena, they have so far eluded experimental verification. Indeed, the progress in the field has been stifled precisely by the lack of experiments, as the scales where quantum gravitational effects are assumed to take place are far beyond current technology.
 
An alternative research program has put forward proposals aimed at finding indirect traces highlighting the overall nature of gravity while still being accessible to tabletop experiments~\cite{carney2019tabletop}. Examples are gravitational decoherence~\cite{anastopoulos2013master,blencowe2013effective} and noise induced by gravitons in interferometers~\cite{parikh2020noise,cho2022quantum}.
Among these alternatives, one of the most promising experimental proposals  is  based on the gravitationally induced entanglement (GIE), which would be a clear indication of the quantumness of gravity~\cite{bose2017spin,marletto2017gravitationally,krisnanda2020observable,van2020quantum,matsumura2020gravity,yi2022spatial}. The idea, borrowed from quantum information theory, is that if gravity is able to mediate quantum correlations between otherwise isolated quantum systems, then it cannot be classical. More specifically, the channel it determines cannot be described in terms of Local Operations and Classical Communication (LOCC). Indeed LOCCs capture effectively the expectation one has about what a classical interaction between quantum systems should have to operate and are known for being entanglement non-increasing~\cite{horodecki2009quantum}.

Parallel to both these efforts, another bold alternative has emerged: the development of hybrid models of classical-quantum interaction~\cite{diosi2014hybrid,diosi2024classical}. These mathematical frameworks are able to describe a classical system interacting with a quantum one in a manner free of inconsistencies. The most immediate application of these has been the development of many models of classical gravity interacting with quantum matter  \cite{tilloy2016sourcing,tilloy2018ghirardi,oppenheim2023postquantum,layton2023weak} to contrast their prediction with the usual quantum gravity models. 

These models have been shown to have an interesting feature: in order to describe a classical system (e.g. potentially the gravitational field) interacting with 
a quantum one, the former must be endowed with some stochasticity~\cite{diosi2023hybrid,oppenheim2023postquantum} and, moreover, the full evolution is necessarily irreversible~\cite{oppenheim2023gravitationally}. To describe such irreversible evolution, the equations underpinning these models share the same structure of the familiar Markovian, completely positive master equations, which are common in the open quantum system literature.

Surprisingly, these hybrid models, while keeping gravity described by a classical field, have been shown to not necessarily be LOCCs~\cite{tilloy2017principle}. Therefore, some entanglement generation \emph{could} take place within those descriptions. The analysis of their entangling power thus becomes essential within the current experimental effort to test the quantization of gravity via GIE. Indeed, if also hybrid models of gravity can generate entanglement, the conclusions one could draw on the nature of gravity --- assuming some degree of entanglement is measured --- could be questioned.  

The aim of the work is  to analyze the entangling power of simple hybrid models in the regimes where the GIE experiments are set to take place. In the preliminary Section~\ref{Sec:Ent}, we discuss more broadly the  entanglement generation in a particular class of Markovian, Gaussian bipartite master equations, as this mathematical framework also encompasses a vast class of hybrid models in appropriate regimes. In Section~\ref{Sec:Hyb}, we apply the developed tools  to a toy model and to the Tilloy-Di\'osi model~\cite{tilloy2016sourcing}. We prove that such models are indeed capable of generating entanglement and find the root of this phenomenon in a certain degree of non-locality of the theories, which we characterize. 
Complementarily, we study the amount of entanglement generated in the context of the experimental proposal presented in Ref.~\cite{krisnanda2020observable}. We find that GIE experiments would be able to falsify or strongly constrain the Tilloy-Di\'osi~\cite{tilloy2016sourcing} and the Di\'osi-Penrose models~\cite{diosi1989models,penrose1996gravity}  depending on the amount of measured entanglement. We draw our conclusions in Section~\ref{sec.conc}.

\section{Entanglement in Markovian Master Equations}\label{Sec:Ent}
The systems under scrutiny in GIE experiments are the centers of mass of two massive particles, which we here describe by dimensionless quadrature operators $\hat r = (\hat q_1,\hat p_1,\hat q_2,\hat p_2)$ obeying canonical commutation relations $\comm{\hat r_i}{\hat r_j} = i \Omega_{ij}$, where $\Omega$ is the symplectic matrix; see Appendix~\ref{APP_Tools} for further details.  We will restrict our attention to a particular class of dynamical evolutions, which are described by the Gorini-Kossakowki-Sudharshan-Lindblad (GKSL) generator  with a quadratic Hamiltonian and Lindblad operators given by $\hat r$. The corresponding master equation reads
\begin{equation}\label{MAIN_EQ:GKLSGauss}
    \partial_t\hat\rho_t = -i\comm{\frac{1}{2}\hat r_i H_{ij}\hat r_{j}}{\hat \rho_t} -\kappa_{ij}\Big[\hat r_i\hat\rho(t)\hat r_j - \frac{1}{2}\acomm{\hat r_j\hat r_i}{\hat\rho(t)}\Big]\,.
\end{equation}
In Eq.~\eqref{MAIN_EQ:GKLSGauss},  the Hamiltonian $H$ is a $4\times4$ real symmetric matrix, while  the Kossakowski matrix $\kappa$  is a $4\times4$ complex matrix (i.e.~$\kappa\in M(4,\mathbb C)$). If $\kappa \ge 0$ then complete positivity of the evolution is guaranteed by the GKSL theorem~\cite{gorini1976completely,lindblad1976generators}. Although complete positivity, as a mathematical tool, will play an major role in the analysis, we only require the evolution~\eqref{MAIN_EQ:GKLSGauss} to be positive. Master equations of this type are known as Gaussian, as they are the most general Markovian dynamics that preserve the Gaussian form of an initially Gaussian state~\cite{ferraro2005gaussian}. While certainly a restrictive set, a promising GIE proposal described in Ref.~\cite{krisnanda2020observable} works within the Gaussian regime. Moreover,  the irreversible contribution is oftentimes an effective description of the aforementioned hybrid models of classical gravity~\cite{kafri2014classical,gaona2021gravitational}, as we will explore in Section~\ref{Sec:Hyb}. Therefore, the form of Eq.~\eqref{MAIN_EQ:GKLSGauss} is general enough to form a solid ground on which to explore both the predictions of hybrid models and describe GIE experiments. 

Thus, we wish to explore, in general fashion, for which forms of $H$ and $\kappa$ an evolution of the type~\eqref{MAIN_EQ:GKLSGauss} is able to generate entanglement in at least one previously unentangled state. This question is rather trivial when $\kappa = 0$, that is, for a unitary evolution. Loosely speaking, the presence of any interaction Hamiltonian between two parties is enough to establish some degree of entanglement. For more general evolutions $\kappa \neq 0$, the issue becomes more subtle as the extra irreversible contributions can prevent the generation of entanglement (or lead to its sudden-death) via decoherence \cite{yu2009sudden,almeida2007environment}, but they could also, potentially, contribute to its formation \cite{benatti2003environment,benatti2006entangling,jakobczyk2002entangling}. 

To study whether~\eqref{MAIN_EQ:GKLSGauss} can map a separable state into an entangled one, we exploit the well-known fact that a  Gaussian state $\hat \rho$ is separable \emph{if and only if} it is positive under partial transposition (PPT), i.e.~$\hat \rho^{\text{PT}}\ge 0$~\cite{simon2000peres}. Therefore, the study of entanglement generation due to the master equation \eqref{MAIN_EQ:GKLSGauss} can be translated to study the positivity of the dynamics of the partially transposed state $\hat \rho^{\text {PT}}$.

We thus construct the equation of motion governing the evolution of the partially transposed state. To render the bipartition more evident, we can write the matrices appearing in \eqref{MAIN_EQ:GKLSGauss} as
\begin{equation}\label{MAIN_EQ:Bipart}
    H = H_\text{loc} +\begin{bmatrix}
        0&h\\
        h^T& 0
    \end{bmatrix}
    \,,\quad
    \kappa = \begin{bmatrix}
        k_1& k_{12}\\
        k_{12}^\dagger & k_2
    \end{bmatrix}\,,
\end{equation} 
where the $4\times 4$ matrix $H_\text{loc}$ collects the terms that are local to the two parties,  the $2\times 2$ matrices $h$ and $k_i$, $k_{12}$ are, respectively, the interaction Hamiltonian and the local and bipartite irreversible contributions to the dynamics. 
The effect of the PT with respect to, say, the second party on the products of $\hat x_2$ or $\hat p_2$ and the density matrix $\hat \rho$ is 
\begin{equation}\label{MAIN_EQ:PTonRho}
    \begin{cases}
    \hat x_2 \hat \rho \overset{\text{PT}}{\longleftrightarrow} \hat \rho^{\text{PT}}\hat x_2\,,\\
    \hat p_2 \hat \rho \overset{\text{PT}}{\longleftrightarrow} -\hat\rho^{\text{PT}}\hat p_2\,,
    \end{cases}
\end{equation}
while leaving unaltered the order of the products of $\hat x_1$ or $\hat p_1$ with $\hat\rho$. By transforming \eqref{MAIN_EQ:GKLSGauss} according to the prescriptions \eqref{MAIN_EQ:PTonRho}, one finds a new GKSL-type of equation involving a new Hamiltonian $H^\text{PT}$ and a new Kossakowski matrix $\kappa^\text{PT}$; the details are reported in Appendix~\ref{APP_B}. For matters of positivity, the former is not relevant, while the crucial object is $\kappa^\text{PT}$, which reads
\begin{equation}\label{MAIN_EQ:KappaPT}
     \kappa^{\text{PT}} = \Theta
    \begin{bmatrix}
        k_1&\Re[k_{12}] +ih\\
        \Re[k_{12}]-ih& k_2^T\\
    \end{bmatrix}\Theta\,,
\end{equation}
where $\Theta = \text{diag}[1,1,-1,1]$. We see that the interaction Hamiltonian $h$ of the original master equation now appears in the irreversible part of the evolution in a manner akin to dissipation: it is this misplacement that can be the source of negative eigenvalues in $\hat \rho^{\text{PT}}$~\cite{benatti2003environment,benatti2006entangling}.

Now, complete positivity of the partially transposed evolution, i.e.~the positivity of the matrix $\kappa^{\text{PT}}$, is a sufficient condition to guarantee that any initially positive $\hat \rho^{\text{PT}}$ remains  such over time~\cite{benatti2003environment}. Thus, on Gaussian states, it can certify whether a certain evolution preserves the separability of an initially unentangled state. Conversely, $\kappa^{\text{PT}}\ngeq 0$ is not, in general, indicative of the presence of entanglement generation. 

We prove, however, a rather technical but fundamental result regarding the relationship between positive and completely positive dynamics. The details of the proof can be found in Appendix~\ref{APP_PvsCP}. We find that in the Gaussian regime, for the class of Kossakowski matrices of the form
\begin{equation}\label{MAIN_EQ:kappa2}
    \kappa = \kappa_2\otimes\begin{bmatrix}
        1 & 0\\
        0& 0
    \end{bmatrix}\quad\text{with}\quad \kappa_2\in M(2,\mathbb{C})\,,
\end{equation}
one has that complete positivity and positivity are equivalent:  to guarantee that \emph{all} valid initial states get mapped to positive operators, it is necessary (and not only sufficient) that $\kappa\ge 0$. 
We apply this fact to the PT dynamics: if $\kappa^{\text{PT}}$ falls into the class of Kossakowski matrices \eqref{MAIN_EQ:kappa2} then, since  positivity and complete positivity become equivalent, $\kappa^{\text{PT}} \ngeq 0$ means that there exists some initial state $\hat \rho^{\text{PT}}\ge 0$ that gets mapped to a non-positive operator, which signals the entanglement generation. To fall in this class, the inspection of \eqref{MAIN_EQ:KappaPT} reveals that two conditions have to be met:
\begin{itemize}
    \item[\textit{1)}] the real part of $\kappa$ must be of the form \eqref{MAIN_EQ:kappa2};
    \item[\textit{2)}] the entangling Hamiltonian $h$ must involve positions only.
\end{itemize}
Explicitly, $\kappa = \kappa_2\otimes \left[\begin{smallmatrix}
    1&0\\
    0&0
\end{smallmatrix}\right] $ and with $h = c \left[\begin{smallmatrix}
    1&0\\
    0&0
\end{smallmatrix}\right]$  with $c\in \mathbb{R}$. To check whether entanglement is generated or not, it suffices to evaluate the positivity of the $2\times 2$ sub-matrix $\Re\kappa_2 + c \left[\begin{smallmatrix}
    0&1\\
    1&0
\end{smallmatrix}\right]$ . Hence the evolution is able to generate entanglement \emph{iff}
\begin{equation}\label{MAIN_EQ:EntIFF}
    \det \Re\kappa_2 \ngeq c^2\,,
\end{equation}
which was a result first found in Refs.~\cite{kafri2014bounds,kafri2014classical}. The present result can be extended beyond the two mode case to more general bipartitions since, as proven in Appendix~\ref{APP_Proof}, the equivalence between positivity and complete positivity for matrices~\eqref{MAIN_EQ:kappa2} is valid for general $N$-modes systems. Therefore, the logic applies in all cases in which PPT is a necessary and sufficient condition for separability of Gaussian states and  $\kappa^{\text{PT}}$ is of the $2N$-dimensional generalization of the structure \eqref{MAIN_EQ:kappa2}. Examples can be $1\times N$-modes case or the $M\times N$-modes symmetric case \cite{lami2018gaussian}. 

Notice, as a corollary, that if there is no interaction $c\rightarrow 0$, then the condition \eqref{MAIN_EQ:EntIFF} is never satisfied due to  the positivity of $\kappa$, regardless of its details. Therefore, spontaneous collapse models like the Continuous Spontaneous Localisation (CSL) model \cite{ghirardi1990markov,bassi2003dynamical}, in the regimes where they admit a linearized form \cite{ferialdi2020continuous} like~\eqref{MAIN_EQ:GKLSGauss}, cannot generate entanglement just from the noisy dynamics, a conclusion that is expected but not trivial \cite{benatti2003environment,benatti2006entangling}. In the next section we discuss two examples in which this criterion of entanglement generation, stemming from this instance of equivalence of positivity and complete positivity of the PT dynamics, can be used in the context of classical theories of gravity within GIE experiments. 

\section{Entanglement in models of classical gravity}\label{Sec:Hyb}
Hybrid models coupling quantum systems with classical ones have received renewed attention over the past years~\cite{oppenheim2023postquantum,diosi2023hybrid,layton2024healthier,barchielli2023markovian,barchielli2024hybrid}, in particular within the attempt to provide consistent models of classical gravity, an alternative quite underexplored. Care must be paid when constructing such hybrid models. Indeed, quantum mechanics is intrinsically stochastic, and this trait must be taken into account in the interaction with a classical field, such as (potentially) the gravitational one. A crucial result is that, in order to have a consistent theory, a classical gravitational field must also be  stochastic~\cite{oppenheim2023postquantum,diosi2023hybrid} and the overall hybrid dynamics must be irreversible~\cite{oppenheim2023gravitationally}. While many concerns have been raised on the consistency of such models in the relativistic and general relativistic settings~\cite{diosi2024classical,tilloy2024general}, within the Newtonian regime -- the same in which the GIE experiments are set to take place -- their formulation is rather unambiguous. In this regime it is often the case that the stochasticity of the gravitational field manifests in terms of Markovian master equations for the evolution of the quantum systems~\cite{tilloy2016sourcing,tilloy2018ghirardi}. Such master equations, in appropriate regimes, are of the form \eqref{MAIN_EQ:GKLSGauss} with a suitable choice of $H_{ij}$ and $\kappa_{ij}$.
This opens the possibility of applying the tools developed in the previous section for the analysis of the entangling potential of these hybrid models.
With this in mind, we first analyse a na\"ive  toy model of classical gravity, where we show how seemingly reasonable premises on the nature of gravity can lead to entanglement generation. Then, we move on to the analysis of the more realistic model, which was developed by Tilloy and Di\'osi \cite{tilloy2016sourcing}, when applied to a concrete experimental setup \cite{krisnanda2020observable}.
Because we are interested in comparison with experiments, in what follows $\hat x$ and $\hat p$ refer to the dimensionful positions and momenta operators. 

\subsection{A na\"ive model for classical gravity}
We now develop a toy model of a classical gravitational interaction between two masses, $m_1$ and $m_2$. Instead of specifying the details of how the two masses interact with the gravitational field and how the latter backreacts onto them, we pin down the form of the master equation for just the two masses from general requests on the nature of the gravitational interaction and the regimes involved. The fact that a master equation for the two masses alone suffices to capture the gravitational interaction is not an obvious feature, but one expected in the Newtonian regime, where usually the gravitational field itself can be integrated out.

As anticipated previously, guided by general arguments \cite{angeli2025probing,diosi2023hybrid,diosi2014hybrid,layton2024healthier,galley2023any,oppenheim2023gravitationally} and many examples \cite{kafri2014classical,tilloy2016sourcing,tilloy2018ghirardi,oppenheim2023postquantum,layton2023weak}, we make the assumption that, if gravity is kept classical, the evolution of the density matrix of two gravitationally interacting masses, in the non-relativistic limit, should be described by a Markovian, completely positive master equation. Markovianity is reasonably expected to hold in the Newtonian limit, while complete positivity is necessary for physical consistency~\cite{benatti2005open,oppenheim2023postquantum}. Thus, one has to identify the Hamiltonian and the form of the operators in the extra, irreversible contributions. 

Now, at the non-relativistic level, the gravitational Hamiltonian is expected to be described by a pairwise interaction given by the quantized version of the Newtonian potential $\hat V_\text{Newton} = -G\frac{m_1m_2}{|\hat x_1-\hat x_2|}$, where $G$ is the gravitational constant \cite{colella1975observation}. One of the key properties of the potential is that it depends only on the relative coordinate and, as such, it is translationally invariant. If we require \emph{both} of these properties to be shared also by the hybrid theory, we end up with a master equation for a pair of particles of the type
\begin{equation}\label{MAIN_EQ:MENaive}
    \begin{aligned}
    \partial_t \hat \rho_t =& -\frac{i}{\hbar}\comm{\hat H_0+\hat V_\text{Newton}}{\hat \rho_t}\\
    &- \frac{1}{2\hbar}\comm{f(\hat x_1- \hat x_2)}{\comm{f(\hat x_1- \hat x_2)}{\hat\rho_t}}\,,
\end{aligned}
\end{equation}
where $\hat H_0$ is the free Hamiltonian of the two particles and $f$ is a generic smooth real function with some dependence on $G$. Note that the last term of the master equation does not affect the averages of the positions and momenta so that, when the state is effectively classical (e.g.~a highly populated coherent state), Newton's law is obeyed. 
If we assume that the pair of masses are sufficiently far apart, {i.e.}~at a big average distance $d$  with respect to their displacement during the motion, which we characterize by the spread of the wavefunction of each particle $\Delta x$, we can Taylor expand the Newtonian potential to second order. This gives
    \begin{equation}\label{eq.approx Newton}
        \begin{aligned}
    \hat V_\text{Newton} \overset{d\gg \Delta x}{\rightarrow}&-\frac{Gm_1m_2}{d^{2}}(\hat{x}_{1}-\hat{x}_{2})-\frac{Gm_1m_2}{d^{3}}(\hat{x}_{1}-\hat{x}_{2})^{2},\\
    &\coloneqq K\hat x_1 \hat x_2 + \text{local terms}\,,
\end{aligned}
    \end{equation}
where we have neglected an irrelevant constant term and defined $K = 2Gm^2/d^3$. Now, assuming that, within the same conditions, a linearization can be applied to $f$ as well, we obtain a master equation
\begin{equation}\label{MAIN_EQ:NaiveME}
    \partial_t\hat\rho_t = -\frac{i}{\hbar}\comm{\hat H_0' +K\hat x_1\hat x_2}{\hat\rho_t} -\frac{1}{2\hbar}\kappa^\text{na\"ive}_{ij}\comm{\hat x_i}{\comm{\hat x_j}{\hat\rho_t}}\,,
\end{equation}
where $\hat H_0'$ collects also all the local terms and 
\begin{equation}\label{MAIN_EQ:knainve}
    \kappa^\text{na\"ive} = f'(d)\begin{bmatrix}
        1&-1\\
        -1&1\\
    \end{bmatrix}
    \otimes
    \begin{bmatrix}
        1&0\\
        0&0\\
    \end{bmatrix}.
\end{equation} 
Notably, $\kappa^\text{na\"ive}$ falls into the class introduced \eqref{MAIN_EQ:kappa2}. Since $\kappa^\text{na\"ive}$ is real, this covers  \textit{condition 1)} mentioned in the previous section. Given that the potential too depends only on the positions (\textit{condition 2)}), we can inquire about the generation of entanglement via the criterion \eqref{MAIN_EQ:EntIFF}. We immediately see that 
\begin{equation}
\det\Re\kappa_2^\text{na\"ive}= f'(d)^2\det \begin{bmatrix}
1 & -1\\
-1 & 1
\end{bmatrix} = 0< K^2\,,
\end{equation}
so some entanglement will \emph{always} be generated regardless of the form of $f$ or the value of $d$. 

\begin{figure}[t]
\includegraphics[width=\linewidth]{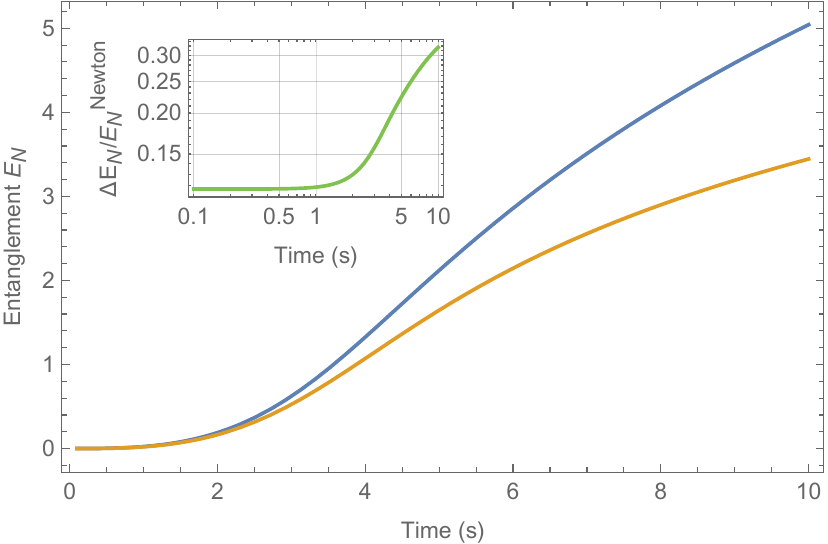}
\caption{Gravitationally induced entanglement quantified by the logarithmic negativity ($E_N$). Blue line: unitary dynamics with the Newtonian potential alone. Orange line: na\"ive model \eqref{MAIN_EQ:NaiveME} with $f = \sqrt{-V_\text{Newton}}$. Inset: relative difference of logarithmic negativity $\Delta E_N /E_N^\text{Newton} = (E_N^\text{Newton} - E_{N}^{\text{na\"ive}})/E_N^\text{Newton}$. 
The parameters used are those of Ref.~\cite{krisnanda2020observable}: $\omega=10^{5}\,${Hz}, $m = 100\,\mu$g, $d = 0.3\,$mm.}
\label{FIG:naive}
\end{figure}

Now, the criterion we used is a witness of whether entanglement generation takes place. However, it gives no information on the quality of such entanglement or its asymptotic fate, for which a more detailed analysis is needed. As a case study, which will be relevant also later on, we consider two masses initially prepared in the ground state of an harmonic trap of frequency $\omega$ and which are then dropped and left to gravitationally interact as in Ref.~\cite{krisnanda2020observable}. For concreteness, we further make the (arbitrary) choice $f = \sqrt{-V_\text{Newton}}$ and compute the logarithmic negativity $E_N$ \cite{vidal2002computable} predicted by the dynamics \eqref{MAIN_EQ:MENaive}. We report the results  in Fig.~\ref{FIG:naive}. The logarithmic negativity can be computed from the two-particle covariance matrix $V$ as follows 
\begin{equation}
    E_N = -\text{log}_2\sqrt{2\Delta(V) - 2\sqrt{\Delta(V)^2-4\det V}}\,,
\end{equation}
where
\begin{equation}
    V = \begin{bmatrix}
        \alpha & \gamma \\
        \gamma & \beta 
    \end{bmatrix},\quad \Delta(V) = \det\alpha + \det\beta -2\det\gamma\,,
\end{equation}
and exploiting the fact that, with the parameters used in Ref.~\cite{krisnanda2020observable}, we have $K/m\omega^2\ll 1$.
The entanglement generated in the na\"ive model is, as expected, less than that in the purely Hamiltonian case, but non-negligible nonetheless: it takes a relatively long time of $t\sim O(s)$ to build up a relative difference $\Delta E_N/E_N^{\text{Newton}}= (E_N^{\text{Newton}} - E_N^{\text{na\"ive}})/E_N^{\text{Newton}} \sim 10^{-1}$. 

In this model the root of the presence of entanglement is found in the high degree of non-locality implicit in a strict dependence on the relative coordinate, which prevents the equation~\eqref{MAIN_EQ:MENaive} from describing a LOCC process. Notably, while at the level of the Hamiltonian, the dependence on the relative coordinates is necessary to have translational covariance of the theory, this is not true for the irreversible contributions. Thus, a way to keep gravity classical and non-entangling would be to make the irreversible terms local, dropping the dependence on the relative coordinates, as it was done in Ref.~\cite{kafri2014classical}.

However, as we will see in the next section, even the more sophisticated classical gravity theories, like \cite{tilloy2016sourcing} which maintain translational covariance while giving up the dependence on the relative coordinate only, end up having some regimes where entanglement can take place.

\subsection{Tilloy-Di\'osi model}
The essence of the Tilloy-Di\'osi (TD) model is that the gravitational field $\phi$ is kept classical and obeys a Poisson equation, but where the source is the average mass density $\langle\hat m(\textbf{x})\rangle$ of the system endowed with some noise to make it compatible with quantum matter: 
\begin{equation}\label{EQ:POiss}
    \grad^2\phi(\textbf{x}) = 4\pi G(\langle\hat m(\textbf{x})\rangle + \xi_t(\textbf{x}))\,.
\end{equation}
This extra noise field $\xi_t(\bf{x})$ is white in time and can have, in principle, any spatial correlation. There is, however, a privileged choice~\cite{tilloy2017principle} of correlation function that relates it to gravity:
\begin{equation}\label{MAIN_EQ:TDcorr}
    \mathbb{E}[\xi_t(\textbf{x})\xi_s(\textbf{y})] = -\frac{2G}{\hbar}\frac{\delta(t-s)}{|\textbf{x}-\textbf{y}|}\,.
\end{equation}
Equation~\eqref{EQ:POiss} can be integrated with usual Green function techniques, and the action of this stochastically sourced field $\phi$ on some massive quantum system can be computed, albeit this procedure has to be done with care, and we refer to the original paper for the details~\cite{tilloy2016sourcing}. Overall, the interaction mediated by $\phi$ can be captured by the following completely positive master equation \cite{tilloy2016sourcing,gaona2021gravitational}
\begin{equation}\label{MAIN_EQ:DP}
    \begin{aligned}
    \partial_t\hat\rho_t =& -\frac{i}{\hbar}\comm{\hat H_{0} +\hat V_\text{Newton}}{\hat\rho_t} \\
    &-\frac{G}{\hbar}\int \frac{\D^3x\D^3y}{|\textbf{x}-\textbf{y}|}\comm{\hat m(\textbf{x})}{\comm{\hat m(\textbf{y})}{\hat\rho_t}}\,,
\end{aligned}
\end{equation}
where we see that, as in \eqref{MAIN_EQ:MENaive}, the Newtonian potential is complemented with some decoherence, this time related to the mass density operator of the system.
To keep the theory free of inconsistencies, a Gaussian smearing of the mass density over a lengthscale $R_0$ has to be introduced \cite{tilloy2016sourcing,ghirardi1990continuous,smirne2014}:
\begin{equation}\label{MAIN_EQ:Smear}
    \hat m(\textbf{x})\rightarrow \hat m_{R_0}(\textbf{x})\coloneqq \int\!\frac{\D^3y}{(2\pi R_0^2)^{3/2}}e^{-(\textbf{x}-\textbf{y})^2/2R_0^2}\hat m(\textbf{y})\,.
\end{equation}
Notice that \eqref{MAIN_EQ:DP} is nothing but the  Di\'osi-Penrose spontaneous collapse model~\cite{diosi1987universal,penrose1996gravity,smirne2014} for two gravitationally interacting masses and that general hybrid models, under reasonable assumptions, reduce to it as well~\cite{layton2023weak,kryhin2025distinguishable}. 
We want to study the entangling potential of \eqref{MAIN_EQ:DP} within the context of one of the GIE proposals presented in \cite{krisnanda2020observable}. Therein, two $100\,\mu$g osmium spheres are held at an initial distance of $d = 0.3\,$mm and are prepared in the ground states of some traps of frequency $\omega = 10^5\,$Hz; they are then freed and left to interact and, thus, entangle during the drop. In these conditions, $d$ is much larger than the initial spread $\Delta x_0 = \sqrt{{\hbar}/{m\omega}} \approx 10^{-16}\,$m, so the Newtonian potential can be effectively expanded to second order as in the previous paragraph. 
However, since the smearing of the masses~\eqref{MAIN_EQ:Smear} affects the Newtonian potential as well, the coupling constant of the bilinear interaction will change to:
\begin{equation}\label{MAIN_EQ:SmearV}
    \hat V_\text{Newton}^{R_0} \approx  K^\text{TD}\hat x_1\hat x_2+\text{local terms},
\end{equation}
where
\begin{equation}\label{MAIN_EQ:SmearVcoeff}
    K^\text{TD} \coloneqq -K\Big[\frac{(d/R_0)^3}{4\sqrt{\pi}}e^{-\frac{d^2}{4R_0^2}}\Big(4\frac{R_0^2}{d^2}+1\Big)-\text{erf}\Big(\frac{d}{2R_0}\Big)\Big],
\end{equation}
and $K$ is defined as~\eqref{eq.approx Newton}. We see that the coupling gains a dependence on both $d$ and $R_0$.
Using as a benchmark the free evolution  and considering an experimental time $t_\text{exp}\sim 10\,$s, the spread will, at most, reach values of $\Delta x_{t_\text{exp}}\approx 10^{-10}\,$m, so the expansion~\eqref{MAIN_EQ:SmearV} holds throughout the evolution.
Furthermore, whenever $\Delta x_t\ll R_0$, it is also legitimate to expand the Lindblad operators of the master equation~\eqref{MAIN_EQ:DP} to linear order, obtaining an equation of the type~\eqref{MAIN_EQ:NaiveME} but with a Kossakowski matrix $\kappa^\text{TD}$\cite{PhysRevD.104.104027,gaona2021gravitational}:
\begin{equation}\label{MAIN_EQ:TDcoeff}
\begin{aligned}
    \kappa^\text{TD}_{11} = \kappa^\text{TD}_{22} &=K\frac{1}{12\sqrt{\pi}}\Big(\frac{d}{R_0}\Big)^3, \\
    \kappa^\text{TD}_{12} = \kappa^\text{TD}_{21} &{=-K^\text{TD}}.
\end{aligned}
\end{equation}
Thus, as long as the two conditions $\Delta x_t \ll (R_0,d)$ are satisfied and the initial states are Gaussian and unentangled, we can use the criterion \eqref{MAIN_EQ:EntIFF} to check if entanglement will be generated. Explicitly, the evolution is  generating entanglement \emph{iff}
\begin{equation}\label{MAIN_EQ:EntCondTD}
    (\kappa_{11}^\text{TD}/K^\text{TD})^2 < 2.
\end{equation}
Since the coefficients of $\kappa^\text{TD}_{ij}$ are functions of $d$ and $R_0$, this condition sets the relative lengthscale over which entanglement can be generated. Numerically, we infer that entanglement will take place only for $d\lesssim 0.85 R_0$, in agreement with one of the setups considered recently \cite{trillo2024di}. This gives an information-theoretic meaning to $R_0$: it is, roughly, the distance over which this ``classical" gravitational field is able to carry quantum information before the intrinsic stochasticity washes it away. 
We emphasise, as it happened in the na\"ive model, that the root of this entanglement generation is some degree of non-locality, encoded in $R_0$. To put it concretely, while these models keep gravity classical and thus allow only classical communication, the information that is being communicated is non-local in nature, breaking the LOCC theorem.
In any case, given the above, the simple detection of GIE would not in principle falsify the TD model, and a more detailed analysis is necessary.

\begin{figure}[t]
\includegraphics[width=\linewidth]{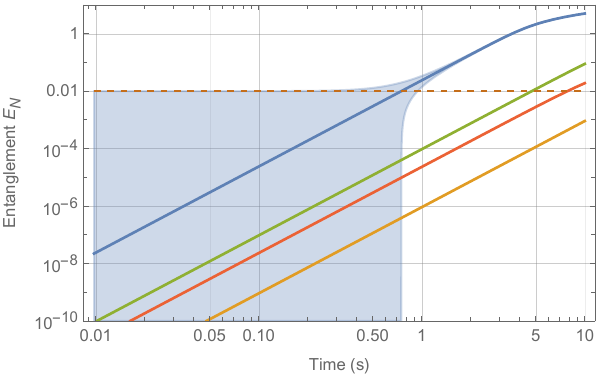}
\includegraphics[width=\linewidth]{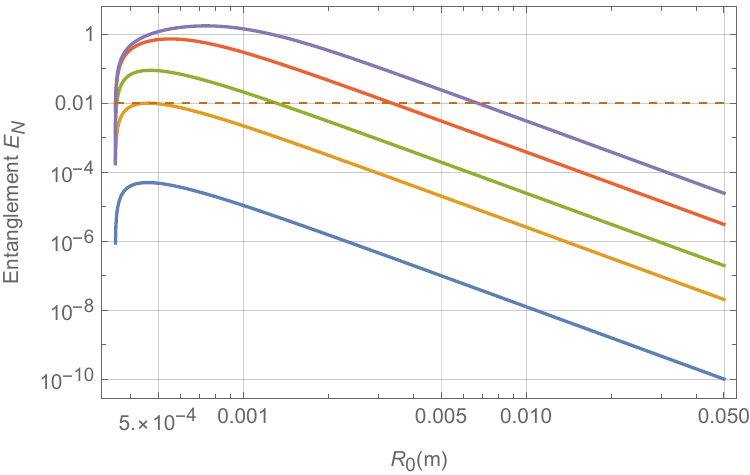}
\caption{Gravitationally induced entanglement quantified by the logarithmic negativity $E_N$ compared to the minimal value of $E_N$ that can be experimentally measured \cite{palomaki2013entangling} (dashed orange line). \textit{Top panel:} $E_N$ vs time. Blu line - Newtonian potential alone with the corresponding  compatible values within a sensitivity of $10^{-2}$ (shaded area). Orange, green, and red lines - TD model with $R_0 = 3\,$mm ($d/R_0 = 0.1$), $R_0 = 0.5\,$mm ($d/R_0 = 0.6$) and $R_0 = 0.37\,$mm ($d/R_0 = 0.8$). 
\textit{Bottom panel:} $E_N$ vs $R_0$. Different experimental times are considered: 
$t = 0.8$\,s (blue line), $t = 4.7$\,s (orange line) and $t = 10$\,s (green line), $t = 25$\,s (red line), $t = 50$\,s (purple line).
The parameters used are those of Ref.~\cite{krisnanda2020observable}: $\omega=10^{5}\,${Hz}, $m = 100\,\mu$g, $d = 0.3\,$mm.}
\label{FIG:NegComp}
\end{figure}

Notice that the condition set by~\eqref{MAIN_EQ:EntCondTD}, for the distance $d=0.3$mm considered in Ref.~\cite{krisnanda2020observable}, already implies that entanglement generation will take place only for values above $R_0 \simeq 3.5\times 10^{-4}$\,m.
We have evaluated the logarithmic negativity for the setup presented in Ref.~\cite{krisnanda2020observable} with various values of $R_0$ to study explicitly the dynamical evolution of the entanglement.
From the top panel of Fig.~\ref{FIG:NegComp}, we see that the entanglement produced in the TD model is, as expected, substantially smaller than that produced by the Newtonian potential alone. 
Notice that two effects are contributing to this significant reduction: not only decoherence but, more importantly, the smearing of the masses, which severely damps the potential, thus reducing its entangling properties. Indeed, for large values of $R_0$ with respect to $d$ 
the smeared potential in \eqref{MAIN_EQ:SmearVcoeff} scales as $(d/R_0)^3$. 

Assuming a sensibility on the values of $E_N$ of $10^{-2}$ (extrapolated from cavity experiments \cite{palomaki2013entangling}), a GIE being compatible with the Newtonian potential would be detectable after $t\approx 0.8$\,s~\cite{krisnanda2020observable}. Instead, for the TD model,  the value $E_N=10^{-2}$ is never reached for any $R_0$ for an experimental time smaller than $t \approx 4.7$\,s. A more detailed analysis is shown in the bottom panel of Fig.~\ref{FIG:NegComp}. {Referring again to the top panel, after $t\approx0.8\,$s, the difference between the $E_N^\text{Newton}$ and $E_N^\text{TD}$ is consistently larger than the projected sensitivity of $10^{-2}$ regardless of the value of $R_0$, so the two can be experimentally discriminated. Thus, entanglement detection compatible with Newtonian predictions for the setup~\cite{krisnanda2020observable} would falsify the TD model entirely.}

\begin{figure}[t!]
\includegraphics[width=\linewidth]{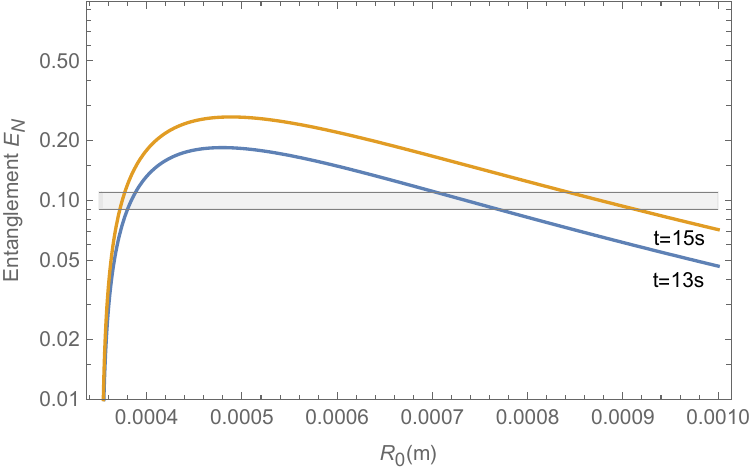}
\caption{Logarithmic negativity ($E_N$) for the TD model for experimental times $t = 13$\,s (blue line) and $t = 15$\,s (orange line) compared against a speculative measured value of $E_N = 0.10\pm 0.01$ (gray shaded region). Only the values of $R_0$ for which the curves intersect the gray band would be compatible with the results. The parameters used are those of Ref.~\cite{krisnanda2020observable}: $\omega=10^{5}\,${Hz}, $m = 100\,\mu$g, $d = 0.3\,$mm.}
\label{FIG:ExpComp}
\end{figure}

If the experiment turns out \emph{not} to agree with Newtonian predictions, then, depending on the measured value of $E_N$, one can constrain the TD model. A speculative example is shown in Fig.~\ref{FIG:ExpComp}.
Whatever the measured value of $E_N$ turns out to be, due to  condition~\eqref{MAIN_EQ:EntCondTD} no value of smaller than $R_0\simeq 1.18d$ (i.e.~$d/R_0\simeq0.85$) can be compatible with entanglement detection. For the distance $d=0.3\,$mm considered in~\cite{krisnanda2020observable}, this would set the constraint $R_0\gtrsim 3.5\times 10^{-4}\,$m, which is six orders of magnitude above the current best experimental bound on the Di\'osi-Penrose model, which is $R_0\geq 4\times 10^{-10}$\,m~\cite{arnquist2022search}. However, we point out that to estimate this improvement accurately, a more realistic analysis including also other decoherence sources is necessary.

In the  case of no entanglement being detected at all, only narrow intervals of $R_0$ would be excluded, since both values below the critical $3.5\times 10^{-4}$\,m  as well as very large values (above a value of $R_0$ set by the runtime of the experiment, for example $R_0\approx3.4\times 10^{-3}$\,m for the red line in the bottom panel of~Fig.~\ref{FIG:NegComp}) would be compatible with this result. Note that values of $R_0\leq3.5\times 10^{-4}$\,m  would be compatible with the experiment thanks to the strength of decoherence, while the very large values of $R_0$ (e.g., $R_0\geq3.4\times 10^{-3}$\,m for the red line in the bottom panel of~Fig.~\ref{FIG:NegComp}) would be so due to  the weakness of the smeared potential.

Finally, let us note that the amount of TD entanglement in Fig.~\ref{FIG:NegComp},  while being small, is still substantially more than that found in Ref.~\cite{trillo2024di}. The reason for this discrepancy is rooted in the specifics of the setups therein analyzed \cite{bose2017spin,marletto2017gravitationally,van2020quantum}, where two massive systems at a distance $d$ need to each be in a cat state delocalized over roughly $\Delta x\sim d$. For entanglement to be generated in these setups, one needs $d \sim R_0 \sim \Delta x$, while in Ref.~\cite{krisnanda2020observable} one has $d\sim R_0 \gg \Delta x$. The smearing $R_0$ being much bigger than the spread of the wavefunction of the two particles renders the decohering dynamics much less effective with respect to the experiment with highly delocalized masses. Or, which is the same, the coherence generated with localized masses is more robust to the noise than the case with delocalized ones for a given $R_0$. Thus, structurally, the setup with delocalized states seems more effective to rule out values of $R_0$ much larger than $d$, although it is much more technologically challenging with respect to the proposal in 
Ref.~\cite{krisnanda2020observable}.

\section{Conclusion}\label{sec.conc}
In this work we have developed tools to study entanglement generation in Markovian, Gaussian master equations describing hybrid classical-quantum theories of gravity. In particular, a class of $N$-modes continuous variable Markovian master equations for which complete positivity and positivity become mathematically equivalent was identified. Many physically relevant examples fall within this class. By hinging on the close relation between positive maps and entanglement, a simple criterion to assess the entangling properties of a large class of Markovian, Gaussian maps is drawn.

We have applied these results  within the current experimental efforts to detect gravitationally induced entanglement. We have proven that hybrid models can indeed predict entanglement generation and have shown explicitly that the root of this behavior is in some degree of non-locality implicit in the theory. 
Our detailed analysis on the TD model, in qualitative agreement with complementary results of Ref.~\cite{trillo2024di}, identified in the parameter $R_0$ the lengthscale over which quantum information can be carried by the classical gravitational field. 
Nevertheless, the quantitative analysis of the predictions of the TD model has shown that entanglement detection in a setup as that proposed in  Ref.~\cite{krisnanda2020observable}  still has the potential to either falsify or set  strong bounds on $R_0$ for the TD and the Di\'osi-Penrose models.

\section*{Acknowledgments}
{We thank Nicolò Piccione for useful comments on the manuscript}.
We acknowledge the PNRR PE Italian National Quantum Science and Technology Institute (PE0000023), {the EU EIC Pathfinder project QuCoM (101046973)} and the University of Trieste (Microgrant LR 2/2011). 

\hfill
\appendix
\section{Tools for continuous variables}\label{APP_Tools}
The systems we are interested in studying are collections of $N$ modes described by quadrature operators $\hat r = (\hat x_1,\hat p_1,\dots,\hat x_N, \hat p_N)$ which, for simplicity, we assume to be dimensionless. The corresponding Hilbert space is $\mathcal{H} = L^{2}(\mathbb{R}^{N})$ and the state of the $N$-mode system is described by a density matrix $\hat \rho$, which is a positive operator of unit trace acting on $\mathcal{H}$. Among all the possible states, a particularly important set is that of \emph{Gaussian} states \cite{ferraro2005gaussian}, which can be defined as all ground and thermal states $\hat \rho = e^{-\beta\hat H}/\mathcal{Z}$ of quadratic Hamiltonians $\hat H = \frac{1}{2}\hat r^T H\hat r$ with $H$ being a $2N\times 2N$ real, symmetric and positive definite matrix, $\beta$ denoting the inverse temperature and $\mathcal Z$ the corresponding partition function. 

The commutation relations of the quadratures can be compactly described as \begin{equation}\label{MAIN_EQ:CCR}
    [\hat r_i,\hat r_j] = i \Omega_{ij}\,,
\end{equation}
where the symplectic matrix $\Omega$ is defined as
\begin{equation}
\Omega =\oplus_{i=1}^N \omega = \mathbb{I}_N\otimes\omega,\quad \omega = \begin{bmatrix}
0&1\\
-1&0
\end{bmatrix}\,.
\end{equation}

The set of $2N\times 2N$ real matrices $S$ such that $S\Omega S^T = \Omega$ form the symplectic group, which we will denote by $\text{Sp}(2N,\mathbb{R})$. Since $S\Omega S^T = \Omega$, a transformation of the coordinates described by a symplectic matrix is canonical, it preserves the commutation algebra~\eqref{MAIN_EQ:CCR}. The non-commutativity of the modes has important physical consequences, the most dramatic of which is the Heisenberg uncertainty principle: it is impossible to know with arbitrarily high precision the value of two non-commuting observables. By introducing the covariance matrix $V_{ij} \coloneqq \langle\acomm{\hat r_i - \langle\hat r_i\rangle}{\hat r_j - \langle\hat r_j\rangle}\rangle/2$, the uncertainty principle can be formulated in a very compact manner as \cite{ferraro2005gaussian,lami2018gaussian}
\begin{equation}\label{MAIN_EQ:UnPrinciple}
    V +\frac{i}{2}\Omega \ge 0\,.
\end{equation}
This formulation has the further advantage that it is symplectically invariant.  Given a symplectic transformation $\hat r\overset{S^{-1}}{\rightarrow} S^{-1}\hat r$ the covariance matrix will transform $V\overset{S^{-1}}{\rightarrow} S^{-1}V(S^{-1})^T$. The condition \eqref{MAIN_EQ:UnPrinciple} therefore will get mapped as
\begin{equation}
    S^{-1}(V +\frac{i}{2}S\Omega S^{T})(S^{-1})^T\ge 0 \iff V +\frac{i}{2}\Omega\ge 0\,.
\end{equation}
Symplectic matrices further play a special role: to each $S\in\text{Sp}(2N,\mathbb{R})$ can be associated uniquely a unitary operator $\hat U_S$ acting on the Hilbert space $\mathcal{H}$ so that $\hat U_S^\dagger \hat r\hat U_S = S\hat r$. These are called Gaussian unitaries due to the fact that such operators correspond to solutions to the Schr\"odinger equation with quadratic Hamiltonians $\hat U_S =\exp\{-\frac{it}{2}\hat r^T H \hat r\}$ and, as such, they preserve the Gaussian character of an initially Gaussian state. 

\section{Partially transposed master equation}\label{APP_B}
Here we highlight some of the steps to obtain the partially transposed Kossakowski matrix \eqref{MAIN_EQ:KappaPT} of the main text for a bipartite two-mode system. First, we consider the interacting Hamiltonian $\hat H_\text{int} = \frac{1}{2}\hat r^TH_\text{int}\hat r$, where
\begin{equation}
    H_\text{int} = \begin{bmatrix}
        0&h\\
        h^T&0
    \end{bmatrix}\,,
\end{equation}
with $h$ a $2\times 2$ matrix and let us look in particular at the component that couples the positions of the two modes, $h_{11}$. Then, by partially transposing with respect to the second party, one finds:
\begin{equation}
    \begin{aligned}
   -ih_{11}\comm{\hat x_1 \hat x_2}{\hat \rho_t}&\overset{\text{PT}}{\longrightarrow} -ih_{11}\Big(\hat x_1 \hat \rho_t\hat x_2 -\hat x_2 \rho_t \hat x_1\Big)\,,\\
    &=-ih_{11}\Big(\hat x_1 \hat \rho_t \hat x_2-\frac{1}{2}\acomm{\hat x_2\hat x_1}{\hat \rho_t}\Big)\\
    &\quad+ih_{11}\Big(\hat x_2 \hat \rho_t \hat x_1-\frac{1}{2}\acomm{\hat x_1\hat x_2}{\hat \rho_t}\Big)\,,
\end{aligned}
\end{equation}
where in the last step we have simply added and subtracted the anticommutator. We see here explicitly that the interacting Hamiltonian turns into an irreversible, dissipative contribution. The same holds also for bilinear coupling including the momenta, the only difference being that whenever $\hat p_2$ is involved, an extra minus sign is picked up. Overall, the partial transposition of the interacting Hamiltonian generates a contribution to the partially transposed Kossakowski matrix:
\begin{equation}\label{APP_B_EQ:PTHam}
    \kappa_h^{\text{PT}} =\begin{bmatrix}
     .&.&-i h_{11} &ih_{12}\\
     .&.&-i h_{21} &ih_{22}\\
     ih_{11} & -ih_{12} &.&.\\
     ih_{21} & -ih_{22} &.&.\\
    \end{bmatrix}\,.
\end{equation}
Next we study how the partial transposition affects the irreversible contribution to the master equation. Recalling the notation of \eqref{MAIN_EQ:Bipart}, it is clear that $k_1$ will be unaffected since it involves only observables of the first party. Conversely, $k_2$ , which involves only observables of the second one, will maintain the same structure but with the off-diagonal elements interchanged and with an extra minus sign:
\begin{equation}\label{APP_B_EQ:PTkappa2}
    k_2 \overset{\text{PT}}{\longrightarrow} k_{2}^{\text{PT}} = \begin{bmatrix}
        [k_2]_{11} &-[k_2]_{21}\\
        -[k_2]_{12} & [k_2]_{22}
    \end{bmatrix}\,.
\end{equation}
It remains only to evaluate the effects on the irreversible components that mix observables of different modes. Considering again the coefficients that couples only the the positions of the modes
\begin{equation}
    \begin{aligned}
    &[k_{12}]_{11}\Big(\hat x_1\hat \rho_t\hat x_2 - \frac{1}{2}\acomm{\hat x_2\hat x_1}{\hat\rho_t}\Big) + h.c.\\
    &\overset{\text{PT}}{\longrightarrow} [k_{12}]_{11} \Big(\hat x_1\hat x_2 \hat \rho_t - \frac{1}{2}\hat x_1 \hat \rho_t\hat x_2 - \frac{1}{2}\hat x_2 \hat \rho_t\hat x_1) + h.c.\,, \\
    &=-\Re[k_{12}]_{11}\Big(\hat x_1\hat\rho_t\hat x_2 -\frac{1}{2}\acomm{\hat x_2\hat x_1}{\hat\rho_t}\Big)+h.c.\\
    &\quad+i\Im[k_{12}]_{11}\comm{\hat x_1\hat x_2}{\hat\rho_t}\,.
\end{aligned}
\end{equation}
Where we see that now the imaginary parts of the Kossakowski matrix end up behaving in the partially transposed equation as Hamiltonians. 

Collecting all contributions, the partially transposed density matrix still obeys a GKLS-like equation with an interacting Hamiltonian
\begin{equation}
    H^{\text{PT}} = \begin{bmatrix}
        0&h_{\text{PT}}\\
        h_{\text{PT}}^T & 0
    \end{bmatrix}
    ,\quad
    h_{\text{PT}} = 
    \begin{bmatrix}
        -\Im[k_{12}]_{11} & \Im[k_{12}]_{12}\\
        -\Im[k_{12}]_{12} & \Im[k_{12}]_{22}\\
    \end{bmatrix}\,,
\end{equation}
and a partially transposed Kossakowski matrix where $k_{1}^{\text{PT}} = k_1$, $k_{2}^{\text{PT}}$ is given by \eqref{APP_B_EQ:PTkappa2}, and the off-block-diagonal contribution is given by
\begin{equation}
    k^{\text{PT}}_{12} = \begin{bmatrix}
        -\Re[k_{12}]_{11}-ih_{11} & \Re[k_{12}]_{12} + ih_{12}\\
        -\Re[k_{12}]_{12} - ih_{21} & \Re[k_{12}]_{22} + ih_{22}
    \end{bmatrix}\,.
\end{equation}
By introducing the matrix $\Theta = \mathbb{I}_2\oplus\left[\begin{smallmatrix}
    -1&0\\
    0&1
\end{smallmatrix}\right]$, with some algebra, one finally arrives at
\begin{equation}
    \begin{cases}
        H^{\text{PT}} = \Theta\begin{bmatrix}
            0 & \Im k_{12}\\
            \Im k_{12}^\dagger & 0 
        \end{bmatrix}
        \Theta\,,\\
        \kappa^{\text{PT}} =\Theta
    \begin{bmatrix}
        k_1&\Re[k_{12}] +ih\\
        \Re[k_{12}]-ih& k_2^T\\
    \end{bmatrix}\Theta\,,
    \end{cases}
\end{equation}
as reported in the main text.

\section{Positivity and Complete Positivity}\label{APP_PvsCP}
Under very general conditions, the time evolution of the density matrix $\hat \rho$ is described by a master equation of the following form \cite{gorini1976completely}:
\begin{equation}\label{MAIN_EQ:GKLS}
    \partial_t\hat\rho(t) = -i\comm{\hat H }{\hat \rho(t)} -\kappa_{ij}\Big[\hat L_i\hat\rho(t)\hat L_j^\dagger - \frac{1}{2}\acomm{\hat L_j^\dagger\hat L_i}{\hat\rho(t)}\Big]\,,
\end{equation}
where $\kappa_{ij}$ are the elements of a Hermitian $2N\times2N$ matrix $\kappa$, known as the Kossakowski matrix, and $\hat L_i$ are generic operators on $\mathcal{H}$. In general, a master equation as in \eqref{MAIN_EQ:GKLS} can be constructed from a microscopic model describing a system interacting with an environment after having traced out the degrees of freedom of the latter~\cite{hu1992quantum}. Alternatively, one can employ a master equation of this form as a tool to effectively describe the system-environment interaction but without a specific underlying microscopic model~\footnote{See Sec. II of~\cite{alicki2002invitation}}, or even postulate it as a physically consistent fundamental law of dynamical evolution, as it happens in spontaneous collapse models~\cite{bassi2003dynamical}.
A key feature of any dynamical evolution of the type \eqref{MAIN_EQ:GKLS} is that, whatever the choice of $\hat{L}$, it must preserve the statistical interpretation of $\hat \rho$. Therefore, the dynamical map $\Lambda_t$, which provides the solution $\hat \rho(t)=\Lambda_t \hat \rho(0)$ of \eqref{MAIN_EQ:GKLS}, must map any positive operator to a positive operator. The characterization of positive maps is a notoriously difficult problem \cite{gurvits2004classical}, however within this class of (Markovian) evolutions of continuous variable (CV) systems, the issue can be posed in a relatively simple language. In fact, any dynamical evolution is constrained to satisfy Heisenberg's uncertainty principle \eqref{MAIN_EQ:UnPrinciple}.
Any dynamics that induces, starting from a \emph{bona fide} initial state, a violation of this condition is unphysical. Since the evolution \eqref{MAIN_EQ:GKLS} structurally preserves both the trace and the hermiticity of $ \hat\rho$, the culprit for a violation of \eqref{MAIN_EQ:UnPrinciple} must be the lack of positivity. 

Positivity, while being a fundamental requirement,  is not sufficient to guarantee the physicality of the evolution due to the existence of \emph{entanglement}. A system could be, in principle, always coupled to a completely inert 
ancilla, say an $n$-level system. Therefore, not only $\Lambda_t$ but also the map $\Lambda_t\otimes \mathbb{I}_n$ should be positive to guarantee the physicality of the evolution, and this should hold true for any $n$ and for any state $\hat \rho$. A map satisfying such a condition is called completely positive \cite{benatti2005open} and it is more easily characterized than its $n=1$ counterpart: a map $\Lambda_t$ giving the solutions to~\eqref{MAIN_EQ:GKLS} is completely positive \emph{iff} the matrix $\kappa$ is positive, that is, $\forall z\in \mathbb{C}^{2N}$ we have $ z^\dagger \kappa z \ge 0$ \cite{lindblad1976brownian,lindblad1976generators}.

It is evident that complete positivity implies positivity, while the converse is generally not true, and even in the CV case counter-examples can be constructed \cite{dekker1984fundamental,isar1993density}. However, since the irreversible contributions in~\eqref{MAIN_EQ:GKLS} influence $V$, and thus play a non-trivial role in the uncertainty principle~\eqref{MAIN_EQ:UnPrinciple}, it becomes natural to ask if there exists any condition under which positivity is a sufficient condition for complete positivity, rendering the two features equivalent. To formulate more concretely: are there any Kossakowski matrices $\kappa$ for which respecting the uncertainty principle implies that $\kappa \ge 0$? We restrict our attention to a particular subclass of evolutions that generalize Gaussian unitaries to maps $\Lambda_t$ 
that are solutions to~\eqref{MAIN_EQ:GKLS} that still preserve the Gaussian character of an initial Gaussian state. The master equations whose solutions have this property are those with a quadratic Hamiltonian and in which the $\hat L = \hat r$ \cite{ferraro2005gaussian}, 
so that
\begin{equation}\label{APP_EQ:GKLSGauss}
    \partial_t\hat\rho_t = -i\comm{\frac{1}{2}\hat r_i H_{ij}\hat r_{j}}{\hat \rho_t} -\kappa_{ij}\Big[\hat r_i\hat\rho(t)\hat r_j - \frac{1}{2}\acomm{\hat r_j\hat r_i}{\hat\rho(t)}\Big]\,.
\end{equation}
For this class of equations, given a $N\times N$ Hermitian matrix $\kappa_N$, then for Kossakowski matrices of the form
\begin{equation}\label{MAIN_EQ:Kstructure} 
\kappa = \kappa_N\otimes \tau \quad \text{where}\quad \tau = \begin{bmatrix}
    1&0\\
    0&0
\end{bmatrix}\,,
\end{equation}
the validity of Heisenberg's uncertainty principle indeed guarantees that that $\kappa \ge 0$. The matrix in \eqref{MAIN_EQ:Kstructure} describes a \emph{diffusive} dynamics in which the irreversible contributions have the position operators of the $N$ modes appearing, but not the momenta.  

While the proof can be found in Appendix~\ref{APP_Proof}, here we provide the physical intuition on why this holds true. For $N=1$, $\kappa_N$ reduces to a real number, and the Kossakwski matrix \eqref{MAIN_EQ:Kstructure} simply describes a particle undergoing decoherence in the position basis 
\begin{equation}\label{MAIN_EQ:DecoEx}
    \partial_t\hat\rho_t = -i\comm{\hat H}{\hat\rho_t} -\frac{\kappa_1}{2}\comm{\hat x}{\comm{\hat x}{\hat\rho_t}}\,,
\end{equation}
which could correspond, for example, to the motion of a heavy particle immersed in a gas of smaller particles, where the recoil is neglected \cite{gallis1990environmental,joos1985emergence}. Decoherence in position leads to diffusion in momentum, that is, to a change of $\Delta p^2 = \langle\hat p^2-\langle\hat p\rangle^2\rangle$ with a rate proportional to the coefficient $\kappa_{1}$: for a negative value of $\kappa_1$ one would eventually reach an unphysical state of negative variance. From this argument it can be seen that $\hat x$  plays no special role, but rather it is complementarity that is the crux. The same argument would work, for example, with decoherence in the $\hat p$ basis, and the variance $\Delta x^2$ becoming negative. This indicates that there is a larger class than that described by \eqref{MAIN_EQ:Kstructure} for which the equivalence of positivity and complete positivity holds.

In fact, the proof of the equivalence is based on the symplectically invariant form of the uncertainty principle \eqref{MAIN_EQ:UnPrinciple}, it remains valid for all Kossakowski matrices that can be obtained from $\kappa$ via a symplectic transformation $S^{-1}\in \text{Sp}(2N,\mathbb{R})$, namely
\begin{equation}\label{MAIN_EQ:KStructureS}
    \kappa^{S^{-1}} = S^T\kappa_N\otimes\tau S\,.
\end{equation}
As an example, for $N=1$ if we pick $S = \left[\begin{smallmatrix}
    1&a\\
    0&1
\end{smallmatrix}\right]$, which represents the transformation $\hat x \rightarrow \hat x + a \hat p$, we obtain a master equation of the form
\begin{equation}
    \begin{aligned}
    \partial_t\hat \rho_t =& -i\comm{\hat H }{\hat\rho_t} - \kappa_{1}\comm{\hat x}{\comm{\hat x}{\hat\rho_t}} - \kappa_1a^2\comm{\hat p}{\comm{\hat p}{\hat \rho_t}}\\
    &-2\kappa_1a\comm{\hat x}{\comm{\hat p}{\hat \rho_t}}\,,
\end{aligned}
\end{equation}
which (apparently) describes a richer diffusive structure and for which the equivalence is not as evident as for \eqref{MAIN_EQ:DecoEx}, but it holds nevertheless as the two are related by $S^{-1}$. 

In the class of master equations corresponding to the form \eqref{MAIN_EQ:Kstructure}, where the irreversible part of the evolution involves only the positions of the particles, there are many physically interesting evolutions. Examples are given by certain regimes of the aforementioned collisional models for decoherence \cite{joos1985emergence,gallis1990environmental}, collapse models \cite{bassi2005collapse,diosi1989models,ferialdi2020continuous} and, which will be of interest later on, of hybrid classical-quantum models for gravity \cite{kafri2014classical,gaona2021gravitational}.

\subsection{Proof}\label{APP_Proof}
The proof is done by contradiction: let us assume that $\kappa \ngeq 0$ so there exists a $2N$ complex vector $w$ for which $w^\dagger \kappa w<0 $. We will show that there exists some valid initial state that evolves to an unphysical state. To do this, we need to evaluate the time evolution of the covariance matrix under \eqref{MAIN_EQ:GKLSGauss}, which is governed by the differential equation
\begin{equation}
    \dot V_t = (\Omega H -\Omega \Im \kappa)V_t -V_t(H \Omega +\Im \kappa\Omega) + \Omega^T\Re\kappa\Omega\,,
\end{equation}
where the dot denotes derivation with respect to $t$ and $(\Re \kappa)_{nm}= (\kappa_{nm} + \kappa_{nm}^*)/2 $, with a similar definition holding for the imaginary part.
Now, let us consider the uncertainty principle \eqref{MAIN_EQ:UnPrinciple} for an arbitrarily short time $\epsilon$, so that we can Taylor expand the condition to linear order
\begin{equation}\label{APP_EQ:UpEps}
    V_0 +\frac{i}{2}\Omega + \epsilon \dot V_t\Big|_0 +O(\epsilon^2)\,.
\end{equation}
Now we choose as a \emph{bona fide} initial state simply $\frac{1}{2}\mathbb{I}_N$. If the modes are a collection of harmonic oscillators, this is nothing but the ground state. This pick is particularly useful because now we can evaluate \eqref{APP_EQ:UpEps} on a vector that belongs to the kernel of $\mathbb{I}_N +i\Omega = \oplus_{n=1}^N \left[\begin{smallmatrix}
    1&i\\
    -i&1
\end{smallmatrix}\right]$. Any vector $u = \oplus_{n=1}^{N}(a_n,ia_n)^T$ with $a_n\in \mathbb{C}$ is a valid choice, and it renders the zeroth-order contribution null. Whether the uncertainty principle is violated or not now rests only upon the sign of
\begin{equation}\label{APP_B_EQ:ThreeContr}
    u^\dagger\dot V\Big|_0u = \frac{1}{2}u^\dagger\comm{\Omega}{H}u -\frac{1}{2}u^\dagger\acomm{\Omega}{\Im\kappa}u +u^\dagger\Omega^T\Re\kappa\Omega u\,.
\end{equation}
Let us consider the last contribution on the right hand side of~\eqref{APP_B_EQ:ThreeContr}
\begin{equation}
\begin{aligned}
    \Omega^T\Re\kappa\Omega &= \Re\kappa_N\otimes\omega^T\tau\omega\,, \\
    &= \Re\kappa_N\otimes\begin{bmatrix}
        0&0\\
        0&1
    \end{bmatrix}\,.
\end{aligned}
\end{equation}
Therefore, evaluating the average over $u$ gives
\begin{align}
    u^\dagger\Omega^T\Re\kappa\Omega u &= \sum_{ij=1}^{N}(\Re\kappa_N)_{nm}(-ia_n^*)(ia_m)\,,\\
    &=a^\dagger\Re\kappa_N a \,.
\end{align}
Next, we can evaluate the second term
\begin{equation}
    \begin{aligned}
    \acomm{\Omega}{\Im\kappa} &= \Im\kappa_N\otimes\acomm{\omega}{\tau}\,,\\
    &=\Im\kappa_N\otimes\omega\,.
\end{aligned}
\end{equation}
Averaging over $u$ gives
\begin{equation}
    \begin{aligned}
    -\frac{1}{2}u^\dagger\acomm{\Omega}{\Im \kappa}u  &= -\frac{1}{2}\sum_{nm=1}^{N}(\Im\kappa_N)_{nm}(2ia_n^*a_m)\,,\\
    &= -ia^\dagger \Im\kappa_Na\,.
\end{aligned}
\end{equation}
Finally, we need to evaluate the Hamiltonian contribution, letting $H = [H_{nm}]$ with $H_{nm}\in \text{Mat}(2,\mathbb{R})$ then
\begin{equation}
    \begin{aligned}
    \comm{\Omega}{H} &= [\omega H_{nm} - H_{nm}\omega] \,,\\
    &= \begin{bmatrix}
        \lambda_{nm}&\mu_{nm}\\
        \mu_{nm} &-\lambda_{nm}
    \end{bmatrix}\,,
\end{aligned}
\end{equation}
where $\lambda_{nm},\mu_{nm}\in\mathbb{R}$. So
\begin{equation}
    \begin{aligned}
    u^\dagger\comm{\Omega}{H}u &= \sum_{nm=1}^{N}\Big(\lambda_{nm}a_n^*a_m + \mu_{nm} ia_n^*a_m + \\
    &\quad +\lambda_{nm}(-ia_n^*)a_m - \mu_{nm}(-ia_n^*)(ia_m)\Big)\,,\\
    &=0\,.
\end{aligned}
\end{equation}
Therefore, the Hamiltonian does not play any role in regard to the positivity of the evolution at short times. Putting everything together, we have
\begin{equation}
    u^\dagger\dot V\Big|_0u = a^\dagger\kappa_N^*a\,,
\end{equation}
with $a$ an \emph{arbitrary} vector on $\mathbb{C}^N$. From here we see why $\kappa$ was chosen with such a specific structure: being a $N\times N$ matrix embedded in a $2N\times2N$ matrix, it is possible to use the $N$ degrees of freedom to kill the zeroth-order term in \eqref{APP_EQ:UpEps}, while leaving $N$ degrees of freedom at one's disposal. The lack of positive definiteness translates into a condition on $\kappa_N$
\begin{equation}
    w^\dagger\kappa w <0 \quad \iff\quad \exists v\in\mathbb{C}^N \quad v^\dagger\kappa_Nv <0\,,
\end{equation}
which we can exploit. By choosing $a = v^*$, we have
\begin{equation}
    u^\dagger\dot V\Big|_0u = a^\dagger \kappa_N^*a \overset{a = v^*}{\rightarrow} (v^\dagger \kappa_N v)^* <0\,.
\end{equation}
This concludes the proof: we have shown that if $\kappa\ngeq0$ then there exists a valid state that, after a time $\epsilon$, evolves to a state violating the uncertainty principle. Therefore, in order to respect \eqref{MAIN_EQ:UnPrinciple} for all valid initial states, it must be that $\kappa \ge 0$.
\bibliography{PosEntGrav.bib}{}
\bibliographystyle{apsrev4-1}

\end{document}